\documentclass[
twocolumn,
showpacs,                     
secnumarabic,
amssymb, 
nobibnotes, 
aps, 
prl]{revtex4-1}

\usepackage{graphicx}
\usepackage{latexsym}
\usepackage{amsmath}
\usepackage{verbatim}

\begin{document}



\title{The Gibbs ``volume'' entropy is incorrect}

\author{Robert H. Swendsen}
\email[]{swendsen@cmu.edu.}
\affiliation{Department of Physics, Carnegie Mellon University, Pittsburgh PA, 15213, USA}

\author{Jian-Sheng Wang}
\email[]{phywjs@nus.edu.sg}
\affiliation{Department of Physics, 
National University of Singapore, Singapore 117551, Republic of Singapore}


\date{23 June 2015}


\begin{abstract}
In  recent papers,
several authors have  
 claimed that 
a definition of
the thermodynamic entropy 
in terms of the logarithm of a volume in phase space,
originally suggested by Gibbs, 
is the only valid definition.
Arguing from the Gibbs entropy,
these authors  claim 
that
thermodynamics cannot be extended to include 
 negative temperatures.
However,
 the Gibbs entropy 
 fails to satisfy the postulates of thermodynamics,
 leading  to serious errors. 
 In particular,
 predictions of the Gibbs entropy 
 for systems with non-monotonic energy densities
are  incorrect.
We show that
the correct expression for the equilibrium entropy 
contains an integral over a surface 
in phase space,
and negative temperature
is a valid thermodynamic concept.
\end{abstract}

\maketitle

\section{Introduction}\label{section: Introduction}

Several  authors have recently
 claimed that 
the only valid definition of
the thermodynamic entropy 
in statistical mechanics
is
in terms of the logarithm of a volume in phase space,
as originally suggested by Gibbs\cite{Gibbs_book,Campisi_SHPMP_2005,DH_Physica_A_2006,Campisi_Kobe_2010,DH_NatPhys_2014, HHD_2014,Campisi_PRE_2015}.
Since the volume entropy 
predicts only positive temperatures,
these authors claim that 
negative temperature 
is not a valid thermodynamic concept.
We will demonstrate that these claims 
are incorrect 
due to the extrapolation of a derivation 
beyond its limits of applicability.
We show that the concept of negative temperature
is a valid extension of  thermodynamics\cite{Frenkel_Warren_2015,swendsen_wang_arxiv_2014,penrose_counter_eg}.

Advocates of 
the volume entropy
have also claimed that it can be used
to calculate the thermal properties 
of very small systems --
 even systems  containing only a single particle.
We believe that this 
is  a misinterpretation of 
thermodynamics,
and a failure to recognize
the importance of fluctuations
in small systems.
We see no advantage to 
using   any form of entropy 
to predict the behavior 
of small systems
over   directly applying 
statistical mechanics.
Thermodynamics predicts measurements 
on macroscopic systems,
so we will only consider systems
that contain many particles.

The 
 debate over 
 the volume entropy
 has also been extended to 
 quantum systems\cite{Campisi_SHPMP_2005,DH_Physica_A_2006,Campisi_Kobe_2010,DH_NatPhys_2014, HHD_2014,Campisi_PRE_2015},
 but
we will limit 
the present discussion  to  
 classical statistical mechanics.
The special features of  quantum entropy 
will be addressed elsewhere\cite{RHS_book,JSW_RHS_unpub}.

We will present a series of 
arguments against the volume entropy.
We will review  the 
postulates of thermodynamics
and show that they are violated by the volume entropy\cite{Tisza,Callen,RHS_book}.
After that,
we will  discuss 
the predictions 
of statistical mechanics for macroscopic systems,
and show how they lead to an expression 
for the equilibrium entropy 
that exactly predicts the mode,
and
which does satisfy the postulates of thermodynamics.
%
We will demonstrate that 
the volume entropy 
makes incorrect predictions  
for systems with non-monotonic 
densities  of states.
Finally,
we will discuss 
the significance of the
the difference 
between the mean and the mode
in macroscopic measurements.
We begin our analysis 
with the role of probability 
in  statistical mechanics and thermodynamics.

\section{Macroscopic experiments and probability}

Thermodynamics 
predicts specific values for quantities 
that can be measured
in experiments on macroscopic systems.
  Statistical mechanics, 
  in contrast,
  predicts a probability distribution,
  %
  where the width of the  distribution 
  gives the thermal fluctuations.
  The two types of predictions are compatible 
  because the probability distributions 
  for macroscopic systems 
  are very narrow, 
  with a relative width of the order of 
  $1/\sqrt{N}$,
  where $N$ is the number of particles.
  If this width is less than the  experimental resolution,
  as it usually is,
  a prediction of statistical mechanics
  is equivalent to a single value.
Since the entropy is a maximum in equilibrium,
it is natural to take this value to be 
the location of the maximum probability,
that is, 
the   mode of the distribution.
  %
  As  shown below,
  this  leads to an expression for the equilibrium entropy
  that correctly predicts the mode in all cases,
  and
  which contains an integral 
  over a surface in phase space.

The  arguments 
that have been advanced in favor of the 
volume entropy 
have been 
based on the  two claims that
(1)
only the mean of the probability distribution 
can represent a thermodynamic prediction,
and 
(2)
 the volume entropy correctly predicts the mean.
The first claim is dubious for reasons given later,
and
all arguments that have been put forward 
for the second
are only valid for systems 
governed by an unbounded energy spectrum.
The predictions of the volume entropy
are very bad for 
non-monotonic densities of states,
as we will demonstrate 
by extending  an argument 
given  in Ref.~\onlinecite{HHD_2014}.

\section{The postulates of thermodynamics}
\label{section: thermodynamics}

We will  discuss why each postulate is needed.

\begin{quote}
\textbf{Postulate 1: Equilibrium states.}
There exist equilibrium states of a composite macroscopic
system that are characterized uniquely by a small number of
extensive variables.
\end{quote}

The extensive variables for systems composed of particles
are  the energies, volumes, and particle numbers 
for each subsystem in the 
(isolated)
composite system.
For simplicity of notation,
we will only consider
a single type of particle,
but the generalization to any  number 
of particle types is trivial.
For other systems,
the magnetization and electrical polarization 
might also be included.

The first postulate 
requires   the equilibrium state of a system
to be independent of its history.
This  leads to  the concept of a state function,
which is a quantity that 
is determined exclusively by 
the equilibrium state of the system,
and therefore
by the values of the extensive variables.

The specification of a composite system
is necessary for the application of the second postulate,
which is a particular form of the second law 
of thermodynamics.

\begin{quote}
\textbf{Postulate 2: Maximization of the entropy.}
For every  isolated composite system,
there exists a state function called the entropy,
such that 
the  equilibrium values assumed by the extensive variables
 in the absence of one or more            
constraints are those that maximize the entropy over the set of
all constrained macroscopic states.
\end{quote}

The word constraint 
refers to any restriction 
on the values of the external variables.
An extreme example 
would occur 
if all subsystems were isolated,
so that all extensive variables
were constant.
The opposite extreme of no constraints
other than the total energy, volume and particle number
would mean everything can be exchanged freely between 
subsystems.

The second  postulate 
enables thermodynamics 
to make testable experimental predictions.
It also indirectly specifies the 
domain 
of thermodynamics.
Because of the existence of fluctuations 
in the absence of a constraint,
prediction is only possible 
if  the resolution of a relevant 
experiment 
 is not sufficient to 
detect  fluctuations.
Since the relative magnitude of fluctuations 
generally goes as 
$1/\sqrt{N}$,
this condition 
is fulfilled by  
most macroscopic experiments.

The first two postulates 
should even be valid for  systems
that  include long-range interactions 
between particles in different subsystems.
The next postulate 
is restricted  
to cases
in which 
direct interactions between 
particles in different subsystems
can be neglected.


\begin{quote}
\textbf{Postulate 3: Additivity.}      
If direct interactions between
particles in different 
 subsystems
are negligible,
the entropy of a composite system 
can be expressed as the sum of the entropies 
of its subsystems.
\end{quote}

If we denote the energy, volume, and particle number of the $j$-th subsystem as
$E_j$,
$V_j$,
and 
$N_j$,
and the set of all energies, volumes, and particles numbers as
$\{\textbf{E}, \textbf{V},  \textbf{N}\}
= \{ E_j,V_j,N_j   \vert j=1, \dots,M \} $,                                             
postulate 3 says that 
 we can write the total entropy of the composite system as
\begin{equation}\label{ST = sum sj}
S_T \left( \textbf{E}, \textbf{V},  \textbf{N} \right)
=
\sum_{j=1}^M
S_j \left(  E_j, V_j, N_j \right)  ,
\end{equation}
where the function $S_j$
is the entropy of the  
subsystem $j$.
%
This postulate
does not distinguish between subsystems
that are isolated from each other 
and subsystems 
that are in thermal contact.
Because entropy is a state function,
the functional dependence of the entropy
on the extensive variables 
cannot 
change.

Since most results in thermodynamics 
are expressed in the language of calculus,
the next postulate plays an important role.


\begin{quote}
\textbf{Postulate 4: Analyticity.}      
The entropy is a continuous and differentiable
function of the extensive variables.
\end{quote}

As an example of why analyticity is important,
consider 
two  thermodynamic 
systems 
in thermal contact
with entropies $S_1$ and $S_2$,
and total entropy 
$S_{1,2}=S_1+S_2$.
In equilibrium,
$S_{1,2}$ 
 takes on its maximum value 
while the total energy
$E_{1,2}=E_1+E_2$, remains constant,
and 
the location of the maximum of $S_{1,2}$  
gives  the  equilibrium values of 
$E_1$ and $E_2$.
$S_1$ and $S_2$
 satisfy  the condition
\begin{equation}\label{d S1+S2 / dE1}
\frac{ \partial  }{ \partial E_1} 
 \left[ 
 S_1(E_1,V_1,N_1)
+
S_2(E_{1,2}-E_1,V_2,N_2)
\right]
=
0      ,
\end{equation}
which can be rewritten as 
\begin{equation}\label{partial E1 = partial E2}
\left( \frac{ \partial  S_1 }{ \partial E_1}  \right)_{V_1,N_1}
=
\left( \frac{ \partial  S_2 }{ \partial E_2}  \right)_{V_2,N_2}   .
\end{equation}

Eq.~(\ref{partial E1 = partial E2})  
exhibits  a property of each system that must  have the same value 
when the systems are in equilibrium.
This means  that if two systems with  the same value of 
$\partial S / \partial E$
are brought into thermal contact,
there will be no net energy transfer,
and no increase in the total entropy.

This immediately gives
the zeroth law of thermodynamics:
If two systems are each in equilibrium with a third system,
they will also be in equilibrium with each other.
These results are also valid for  partial derivatives 
with respect to volume and particle number.

The temperature, $T$, is not 
mentioned in the fundamental postulates\cite{RHS_book,Tisza,Callen,Lieb_Yngvason},
 but it is clear from 
 Eq.~(\ref{partial E1 = partial E2})
 that 
 $\partial S / \partial E$
is  related to the temperature.
If the ideal gas law
 ($PV=N k_B T$)
 is used to define a thermometer,
we can relate the partial derivatives 
in  Eq.~(\ref{partial E1 = partial E2})
to the temperature:
\begin{equation}\label{define T}
\left( \frac{ \partial  S }{ \partial E}  \right)_{V,N}
=
\frac{1}{T}    .
\end{equation}
Without 
 Eq.~(\ref{partial E1 = partial E2}),
there would be no  justification for defining the temperature in 
Eq.~(\ref{define T}).

The analyticity  postulate is essential to the 
standard mathematical manipulations 
in thermodynamics.
%
%
It is, however,
not exactly true for some of the most intensively 
studied models in thermal physics.
To begin with the most obvious,
particle numbers are discrete, not continuous,
but
as a practical matter,
the discreteness is extremely fine grained
and can usually  be ignored with impunity.
%
%
Since the resolution of  thermodynamic measurements
is limited,
the discreteness of $N_j$ is not macroscopically measurable.

The discreteness of  energy eigenvalues in quantum systems
requires a more extensive discussion 
than we have room for in this paper,
and  will be examined elsewhere\cite{RHS_book,JSW_RHS_unpub}.

The next postulate is at the center of the 
debate concerning  the volume entropy.


\begin{quote}
\textbf{Postulate 5: Monotonicity (optional).}      
The entropy is a monotonically increasing function
of the energy for equilibrium values of the energy.
\end{quote}

Although not essential to thermodynamics,
 this postulate 
is usually assumed
in order  to allow the entropy function
of a subsystem
$S = S \left(  E, V, N \right)$
(ignoring subscripts for simplicity)
to be inverted to give 
$E = E \left(  S, V, N \right)$.
%
This inversion  enables  
Legendre transforms to generate the various 
familiar thermodynamic potentials\cite{Callen,RHS_book}.
If non-monotonic entropy functions 
are permitted,
we cannot invert 
the entropy function,
but
we can still use  Massieu functions,
which are Legendre transforms of
$S=S(E,V,N)$\cite{Callen,RHS_book}.
If we define a dimensionless entropy 
$\tilde{S} = S/ k_B$,
its Legendre transform with respect to the variable 
$\beta=1/k_B T$ is 
$\tilde{S}[\beta] = - \beta F \left(T,V,N \right)$,
where $F$ is the Helmholtz free energy.

Because of Eq.~(\ref{define T}),
if the entropy is not a monotonic function of the energy,
the temperature $T=1/k_B \beta$ can be negative.
There is no thermodynamic reason for excluding negative temperatures,
other than the convenience 
of being able to use the more familiar 
Legendre transforms of the energy.


%

\section{Equilibrium statistical mechanics and  entropy}
\label{section: SM}

Fortunately, 
the predictions of statistical mechanics 
are  less subject to debate 
than the definition of entropy.
%
By examining  
how statistical mechanics 
makes  predictions 
for the same experimental questions 
as thermodynamics,
we will derive an expression 
for the equilibrium entropy.
%

We  are concerned with  the properties 
of a composite system containing  $M \ge 2$  subsystems,
%
which is isolated
from the rest of the universe.
The total energy, volume, and particle number(s)
are constant.
%
The phase space 
of subsystem $j$
is
$\{p_j,q_j\}$,
where 
$p_j$ is the set of all momenta 
and 
$q_j$ is the set of all position coordinates.
%
The phase space of the composite system 
is   
$\{p,q\}= \{ p_j,q_j \vert j=1, 2, \dots, M \}$.
We denote 
the  Hamiltonian   
of  subsystem $j$
as 
$H_j( p_j,q_j)$,
and 
assume that we can neglect 
direct interactions between particles in different subsystems.
The  Hamiltonian of the  composite system
can then be written as 
\begin{equation}\label{H total p,q 1}
H_T( p,q)   
=
\sum_j H_j( p_j,q_j) .
\end{equation}
%

%
Consider the microcanonical ensemble 
for the full composite system,
which has
 uniform probability 
in 
$\{p,q\}$,
subject to the fixed values of the total energy, $E_T$,
volume,  $V_T$, 
 and 
 number of particles, $N_T$.
 We can find the probability distribution of
 a macroscopic state
$\{\textbf{E}, \textbf{V}, \textbf{N}\}$
%
by integrating over $p$ and $q$.
\begin{eqnarray}\label{W composite fundamental}
W\left( \textbf{E}, \textbf{V},  \textbf{N} \right)
&=&
\frac{1}{\Omega_T} \,
\frac{N_T!}
{ \prod_j N_j!}
\int dp  \int dq     \nonumber \\
&&
\times
\prod_j \delta \bigl(  E_j - H_j( p_j,q_j) \bigr)    .   
\end{eqnarray}
The constant
$\Omega_T$ is a normalizing factor.
The multinomial factor in
Eq.~(\ref{W composite fundamental}) 
reflects the condition that $N_j$ specifies how many particles 
are in subsystem $j$,
but not which ones;
all permutations of particles are equally probable\cite{RHS_4}.
$W\left( \textbf{E}, \textbf{V},  \textbf{N}\right)$
can also be written as a product of terms, 
with one term from each subsystem,
\begin{equation}\label{W = prod Omega}
W\left( \textbf{E}, \textbf{V},  \textbf{N} \right)
=
\frac{1}{\Omega'_T}
\prod_j  \Omega_j  \left( E_j,V_j,N_j  \right)    ,
\end{equation}
where
\begin{equation}\label{Omega j}
\Omega_j  
=
\frac{1}{h^{3N_j } N_j! }
\int dp_j \, \int dq_j \, 
\delta \bigl(  E_j - H_j( p_j,q_j ) \bigr)   ,
\end{equation}
and the integral over $q_j$ is restricted to the volume 
$V_j$.


If the values of any constrained variables 
are kept constant  in 
$W\left(\textbf{E}, \textbf{V},  \textbf{N} \right)$,
we have
the probability distribution for the 
remaining unconstrained 
 variables.
From this function,
we can predict
both the equilibrium values of the 
extensive variables
and their fluctuations.
$W\left(\textbf{E}, \textbf{V},  \textbf{N}\right)$
will have 
 very narrow peaks around the equilibrium values
 of the unconstrained variables.

As discussed in the Introduction,
we take  
the mode
of $W$
(the location of the maximum probability)
to be  the equilibrium value.
We can then see that the expression
\begin{equation}\label{S of observables 1}
S \left(\textbf{E}, \textbf{V},  \textbf{N} \right)
=
k_B  \ln W\left( \textbf{E}, \textbf{V},  \textbf{N} \right) + {\rm const}	
\end{equation}
satisfies all the postulates of thermodynamics.
By appropriate choice of constants,
we can write the entropy of  a system 
in equilibrium
as 
\begin{equation}
S\left( \textbf{E}, \textbf{V},  \textbf{N} \right)
=
\sum_j S_j\left( E_j,V_j,N_j  \right)  ,
\end{equation}
where the equilibrium entropy of a subsystem is given by
\begin{equation}\label{Sj equilibrium 1}
S_j\left( E_j,V_j,N_j  \right)
=
k_B  \ln \Omega_j  \left( E_j,V_j,N_j  \right)  .
\end{equation}

The expression for the entropy of a subsystem 
in equilibrium 
agrees with what 
is often denoted as
$S_B$,
in honor of   Boltzmann,
who derived the equilibrium energy dependence  
in Eqs.~(\ref{Omega j})  and 
(\ref{Sj equilibrium 1})\cite{Boltzmann,Boltzmann_translation,RHS_4}.
It is also called the surface entropy
because the delta function in 
Eq.~(\ref{Omega j})
limits the integral to a surface in phase space.
However, 
although
Eqs.~(\ref{Omega j}) and (\ref{Sj equilibrium 1}) 
give the equilibrium entropy of a subsystem,
it is 
Eq.~(\ref{S of observables 1})
that actually constitutes 
the definition of the entropy\cite{RHS_1,RHS_4,RHS_5,RHS_unnormalized,RHS_book}.
The distinction
between the  definition of the entropy 
and equations for the equilibrium entropy of a subsystem
 is not essential to the current discussion,
 but it is critical in other contexts.
Liouville's theorem 
requires 
that 
the integral in
Eq.~(\ref{Omega j}) 
must be time-independent
  for irreversible processes
following the release of a constraint.
The  probability distribution 
for the variables
$\{ \textbf{E}, \textbf{V},  \textbf{N} \}$
is  time dependent.
It  will go to 
$W\left( \textbf{E}, \textbf{V},  \textbf{N} \right)$
for irreversible processes,
and the entropy will increase\cite{RHS_1,RHS_6,RHS_unnormalized}.

\section{The volume entropy and the postulates of thermodynamics}
\label{section: volume vs postulates}

 Hilbert, H\"anggi, and Dunkel  (HHD)
have given an extended description 
of how they  interpret the volume entropy\cite{HHD_2014}.
However, 
their interpretation 
 fails to satisfy the first three 
postulates of thermodynamics.
The central problem is that 
they find 
the total entropy of two systems in thermal contact 
 by a new calculation in statistical mechanics,
rather than the sum of the entropies.
This violates  additivity.
Furthermore,
the new entropy for the combined system 
is a function of the total energy,
not the energies of the individual systems.
This eliminates the possibility of predicting 
 the  equilibrium values 
of the individual energies 
from the location of the maximum total entropy.
As far as we can tell,
the volume entropy in this interpretation
leads to no thermodynamic predictions.

Because the total entropy of two systems 
in thermal contact 
differs from the sum of the entropies 
of the individual systems
even if there is no net energy transfer 
due to the thermal contact,
the volume entropy is not
a state function.
Furthermore, the entropy can be changed 
by a measurement,
even if the measurement does not change the 
values of the extensive variables
(see  footnote 24 of Ref.~\cite{HHD_2014}).
Since the volume entropy 
depends on the history of the system,
it is not a state function.
Finally,
in this interpretation,
the volume entropy fails to satisfy 
Eq.~(\ref{partial E1 = partial E2})
($\partial S_1/\partial E_1 = \partial S_2/\partial E_2$).
We regard
Eq.~(\ref{partial E1 = partial E2})
as an essential consequence 
of the postulates of thermodynamics,
but it  is dismissed   as 
``naive''
in Ref.~\onlinecite{HHD_2014}.

There is 
an alternative interpretation 
of the volume entropy 
that would retain additivity.
This interpretation would satisfy postulates 1 and 3.
It does lead to predictions 
for the energies of individual systems
in thermal contact,
but it does not give
the correct equilibrium values  
for 
non-monotonic densities of states, 
as discussed in the next section.

\section{Non-monotonic density of states}
\label{section: T<0}

The arguments in favor of the volume entropy 
are based on
the demonstration in Ref.~\cite{DH_Physica_A_2006}
that it gives the correct mean energy
for systems with unbounded energy spectra,
rather than the mode
 found by our definition in terms of probabilities.
Unfortunately, 
the assumption that the volume entropy would also 
give the  mean for systems 
with bounded energy spectra is not valid. 
An extension of an argument given 
 in Ref.~\onlinecite{HHD_2014},
 shows  that 
 the volume entropy 
 gives incorrect predictions
 for this case.
 This is particularly important
 because 
 these are  the systems 
 at the center of the debate
concerning negative temperatures.

 Consider  two systems that each have  non-monotonic densities of states 
 of the form
\begin{equation}\label{HHD density of states n}
\Omega_j(E_j) \propto E_j^{n_j} (E_{j,{\rm max}} - E_j)^{n_j}  ,
\end{equation}
where
$n_j$ indicates the number of degrees of freedom
(proportional to the number of particles).
The minimum energy is $0$,
and
 $E_{j,{\rm max}}$  specifies the maximum  energy
 in  system $j$.
For the special case of $n_1=n_2=1$
and $E_{1,{\rm max}}= 2E_{2,{\rm max}}$,
Ref.~\onlinecite{HHD_2014}
 investigated  whether 
systems initially at equal temperatures
would have no net energy transfer 
after being brought together. 
 %
Fig.~7 of Ref.~\onlinecite{HHD_2014}
 plots the predictions of the volume entropy and $S_B$ for 
the mean energies of the subsystems at which there would be no mean energy transfer,
and it shows large errors in all predictions.
The authors concluded that
although the predictions of the volume entropy
were incorrect,
so were the predictions of $S_B$.

The weakness of the argument in
Ref.~\onlinecite{HHD_2014}
 is that 
$n_j=1$
corresponds to a  very small system with only a single degree of freedom.
For a macroscopic system,
a large value of $n_j$ would give a more appropriate test.
$S_B$ always predicts the mode correctly,
and the difference between the mode and the mean
goes to zero as $1/n_j$.
If Fig.~7 of Ref.~\onlinecite{HHD_2014}
is redrawn with larger values of  $n_1$ and $n_2$,
agreement of the predictions of 
$S_B$ with the exact mean are excellent.
In contrast,
the predictions of 
the volume entropy 
do not agree with either the mean or the mode.
The volume entropy 
fails this test,
while 
$S_B$
passes.

The failure of the volume entropy
to describe equilibrium  
for systems with a non-monotonic 
density of states 
is due to 
its phase-space integral  
being dominated by  
lower-energy 
states that have no effect on 
 the equilibrium ensemble.
Its failure
shows that it is irrelevant for such systems,
and 
any argument against negative temperatures
based on the volume entropy
is without foundation.

%


\section{Measuring  the mean vs. the mode}

Since the difference between the mean and the mode
plays a role 
in statistical mechanics,
a few comments on their 
relevance for thermodynamics
are in order.

Differences 
between the values of the mean and the mode
 are of order $1/N$,
which makes them  much smaller than the thermal fluctuations,
which are of the order of  $1/\sqrt{N}$.
Furthermore, 
experiments measure the current value of a variable,
which is different 
every time a measurement is repeated
due to the fluctuations.
Even if it were possible to measure a macroscopic variable 
of a system in equilibrium
with perfect accuracy,
the result  would not equal either the mean or the mode,
but would be somewhere in a region of width $1/\sqrt{N}$.
To determine the mean
with a resolution of $1/N$,
we would need at least $N$ measurements,
which would take a long time.
We might imagine making 
an independent measurement  with perfect accuracy 
every second.
Even if  the system had only
$N=10^{12}$ 
particles,
the experiment would still take over $30,000$ years.
%
For $10^{18}$ particles,
it would take  longer than the age of the universe.
Measurement of 
the difference between the mean and the mode  
in a macroscopic system 
is impossible,
even if each individual measurement has no error.

\section{Summary}
\label{section: summary}

We have shown that the Gibbs volume entropy
does  not provide a viable definition of the 
thermodynamic entropy.
In particular,
the failure of the volume entropy
to correctly describe a thermodynamic system
with a non-monotonic density of states 
eliminates the 
argument 
against the concept of negative temperature.
We have given 
an alternative   definition that leads to
 an equilibrium entropy
 that always predicts the mode of extensive variables correctly.
This  definition of the entropy 
satisfies all postulates
and
confirms that
negative temperature 
is a valid extension of
 thermodynamics.

\section*{Acknowledgement}

We would like to thank Roberta L. Klatzky 
for valuable discussions.

\makeatletter
\renewcommand\@biblabel[1]{#1. }
\makeatother

\bibliography{Entropy_citations_1}

\end{document}